\documentclass[12pt]{article}
\usepackage{graphicx}
\def\be{\begin{equation}}
\def\ee{\end{equation}}
\def\bea{\begin{eqnarray}}
\def\eea{\end{eqnarray}}

\catcode`@=12
\begin{document}
\thispagestyle{empty}
\begin{flushright}
UCRHEP-T415\\
June 2006\
\end{flushright}
\vspace{0.5in}
\begin{center}
{\LARGE \bf Quark Masses and Mixing\\ with A$_4$ Family Symmetry\\}
\vspace{1.0in}
{\bf Ernest Ma$^a$, Hideyuki Sawanaka$^b$, and Morimitsu Tanimoto$^c$\\}
\vspace{0.2in}
{\sl $^a$ Physics Department, University of California, Riverside,
California 92521, USA\\}
\vspace{0.1in}
{\sl $^b$ Graduate School of Science and Technology, Niigata University, 
950-2181 Niigata, Japan\\}
\vspace{0.1in}
{\sl $^c$ Department of Physics, Niigata University, 950-2181 Niigata, 
Japan\\}
\vspace{1.0in}
\end{center}
\begin{abstract}\
The successful $A_4$ family symmetry for leptons is applied to quarks, 
motivated by the quark-lepton assignments of SU(5).  Realistic quark 
masses and mixing angles are obtained, in good agreement with data.  
In particular, we find a strong correlation between $|V_{ub}|$ and the 
CP phase $\beta$, thus allowing for a decisive 
future test of this model.
\end{abstract}
 
\newpage
\baselineskip 24pt

Since the original papers \cite{mr01,bmv03} on the application of the 
non-Abelian discrete symmetry $A_4$ to quark and lepton families, 
much progress has been made in understanding the case of tribimaximal 
mixing \cite{hps02} for neutrinos in a number of specific models \cite{a432}. 
As for quarks, the generic prediction \cite{bmv03} is that its mixing 
matrix is just the unit matrix, which can become realistic only if 
small mixing angles can be generated by interactions beyond those of 
the Standard Model, such as in supersymmetry \cite{bdm99}.  Other ideas 
of quark mixing include the judicious addition of terms which break the 
$A_4$ (as well as the residual $Z_3$) symmetry explicitly \cite{a4q}.  
In this paper, motivated by the 
quark-lepton assignments of SU(5), we study a new alternative 
scenario, where realistic quark masses and mixing 
angles are obtained, entirely within the $A_4$ context.

In SU(5) grand unification, the \underline{5}$^*$ representation contains 
the lepton doublet $(\nu,l)$ and the quark singlet $d^c$, whereas the 
\underline{10} representation contains the lepton singlet $l^c$ and 
the quark doublet $(u,d)$ and singlet $u^c$.  In the successful $A_4$ 
model for leptons, $(\nu_i,l_i)$ transform as a \underline{3} 
whereas $l^c_i$ transform as \underline{1}, \underline{1}$'$, 
and \underline{1}$''$.  This means that we should choose
\begin{equation}
d^c_i \sim \underline{3}, ~~~ u^c_i, ~(u_i,d_i) \sim \underline{1}, 
\underline{1}', \underline{1}''.
\end{equation}
Assuming as in the leptonic case three Higgs doublets $\Phi_i = (\phi_i^+,
\phi_i^0)$ transforming as \underline{3} under $A_4$, 
the relevant Yukawa couplings linking $d_i$ with $d^c_j$ are given by
\begin{eqnarray}
&&h_1 d_1 (d^c_1 \phi^0_1 + d^c_2 \phi^0_2 + d^c_3 \phi^0_3) + 
h_2 d_2 (d^c_1 \phi^0_1 + \omega d^c_2 \phi^0_2 + \omega^2 d^c_3 \phi^0_3)
\nonumber\\
&& +~h_3 d_3 (d^c_1 \phi^0_1 + \omega^2 d^c_2\phi^0_2 + \omega d^c_3 \phi^0_3),
\end{eqnarray}
\noindent resulting in the $3 \times 3$ mass matrix:
\begin{equation}
{\cal M}_{d d^c} = \pmatrix{h_1 & 0 & 0 \cr 0 & h_2 & 0 \cr 0 & 0 & h_3} 
\pmatrix{1 & 1 & 1 \cr 1 & \omega & \omega^2 \cr 1 & \omega^2 & \omega} 
\pmatrix{v_1 & 0 & 0 \cr 0 & v_2 & 0 \cr 0 & 0 & v_3},
\end{equation}
where $\omega = \exp(2 \pi i/3)$, $h_i$ are three independent Yukawa 
couplings, and $v_i$ are the vacuum expectation values of $\phi_i^0$. 
(For details of the $A_4$ multiplication rules, see for example 
the original papers \cite{mr01,bmv03} or the more recent review \cite{fuji}.) 
On the other hand, the Higgs doublets linking $u$ with $u^c$ must be 
different because of the latter's $A_4$ assignments.  We choose here two 
Higgs doublets transforming as \underline{1}$'$ and \underline{1}$''$, then
\begin{equation}
{\cal M}_{u u^c} = \pmatrix{0 & \mu_2 & \mu_3 \cr \mu_2 & m_2 & 0 \cr 
\mu_3 & 0 & m_3},
\end{equation}
where $m_2,\mu_3$ come from \underline{1}$'$ and $m_3,\mu_2$ from 
\underline{1}$''$.  This matrix is also symmetric because of the usually 
assumed SU(5) decomposition of $\underline{10} \times \underline{10} \times 
\underline{5} \to \underline{1}$.

In minimal SU(5), there is just one \underline{5} representation of Higgs 
bosons, yielding thus only two invariants, i.e. $\underline{10} \times 
\underline{10} \times \underline{5} \to \underline{1}$ (for the $u u^c$ mass 
matrix) and $\underline{5}^* \times \underline{10} \times \underline{5}^* 
\to \underline{1}$ (for the $l l^c$ and $d^c d$ mass matrices).  The 
second invariant implies $m_\tau = m_b$ at the unification scale which is 
phenomenologically desirable, but also $m_\mu = m_s$ and $m_e = m_d$ which 
are not. To decouple the $ll^c$ and $d^cd$ mass matrices, we follow the 
usual strategy of using both \underline{5}$^*$ and \underline{45} 
representations of Higgs bosons, so that one linear combination 
couples to only leptons, and the other only 
to quarks.  Both transform as \underline{3} under $A_4$.  There are also 
two \underline{5} representations transforming as \underline{1}$'$ and 
\underline{1}$''$ under $A_4$ which couple only to $uu^c$, which must still 
be symmetric.

In the limit $|\mu_2| << |m_2|$ and $|\mu_3| << |m_3|$, we obtain the 
three eigenvalues of ${\cal M}_{u u^c}$ as
\begin{equation}
m_t \simeq |m_3|, ~~~ m_c \simeq |m_2|, ~~~ m_u \simeq \left| {\mu_2^2 \over 
m_2} + {\mu_3^2 \over m_3} \right|
\end{equation}
with mixing angles
\begin{equation}
V_{uc} \simeq {\mu_2 \over m_2}, ~~~ V_{ut} \simeq {\mu_3 \over m_3}, 
~~~ V_{ct} \simeq 0.
\end{equation}

In the $d$ sector, we note first that
\begin{equation}
{\cal M}_{d d^c} {\cal M}_{d d^c}^\dagger = \pmatrix{A |h_1|^2 & 
B^* h_1 h_2^* &  B h_1 h_3^* \cr  B h_1^* h_2 &  A |h_2|^2 &  B^* h_2 h_3^* 
\cr  B^* h_1^* h_3 &  B h_2^* h_3 &  A |h_3|^2},
\end{equation}
where
\begin{equation}
A = |v_1|^2 + |v_2|^2 + |v_3|^2, ~~~ B = |v_1|^2 + \omega |v_2|^2 + 
\omega^2 |v_3|^2.
\end{equation}
Its eigenvalue equation is
\begin{eqnarray}
&& \lambda^3 - \lambda^2 (|v_1|^2+|v_2|^2+|v_3|^2)(|h_1|^2+|h_2|^2+|h_3|^2) 
- 27 |v_1|^2 |v_2|^2 |v_3|^2 |h_1|^2 |h_2|^2 |h_3|^2 \nonumber \\ 
&& + ~ 3 \lambda (|v_1|^2 |v_2|^2 + |v_1|^2 |v_3|^2 + |v_2|^2 |v_3|^2) 
(|h_1|^2 |h_2|^2 + |h_1|^2 |h_3|^2 + |h_2|^2 |h_3|^2) = 0.
\end{eqnarray}

If $|v_1|=|v_2|=|v_3|=|v|$ as assumed in the original papers and all those 
of Ref.~\cite{a432}, then $A = 3|v|^2$, $B=0$, and the three eigenvalues 
are simply $3 |h_{1,2,3}|^2 |v|^2$.  We choose them instead to be different, 
but we still assume $|h_3|^2 >> |h_2|^2 >> |h_1|^2$. In that case, we find
\begin{eqnarray}
m_b^2 &\simeq& (|v_1|^2+|v_2|^2+|v_3|^2)|h_3|^2, \\ 
m_s^2 &\simeq& {3(|v_1|^2|v_2|^2+|v_1|^2|v_3|^2+|v_2|^2|v_3|^2)|h_2|^2 
\over |v_1|^2+|v_2|^2+|v_3|^2}, \\ 
m_d^2 &\simeq& {9 |v_1|^2 |v_2|^2 |v_3|^2 |h_1|^2 \over |v_1|^2|v_2|^2+
|v_1|^2|v_3|^2+|v_2|^2|v_3|^2},
\end{eqnarray}
and the mixing angles are given by
\begin{eqnarray}
V_{sb} &\simeq& \left( {B^* \over A} \right) {h_2 \over h_3}, \\ 
V_{db} &\simeq& \left( {B \over A} \right) {h_1 \over h_3}, \\ 
V_{ds} &\simeq& \left( {AB^*-B^2 \over A^2-|B|^2} \right) {h_1 \over h_2},
\end{eqnarray}
thereby requiring the condition
\begin{equation}
\left| {V_{ds} V_{sb} \over V_{db}} \right| \simeq \left| {A B^* - B^2 \over 
A^2 - |B|^2} \right|,
\end{equation}
which has unity as an upper bound.  Using current experimental values for 
the left-hand side, we see that quark mixing in the 
$d$ sector alone cannot explain the observed $V_{CKM}$.  Taking into 
account $V_u$, we then have $V_{CKM}=V_u^\dagger V_d$.  Hence
\begin{eqnarray}
V_{us} &\simeq& V_{ds} - V_{uc}, \\ 
V_{cb} &\simeq& V_{sb}, \\ 
V_{ub} &\simeq& V_{db} - V_{uc} V_{sb} - V_{ut}.
\end{eqnarray}
We show in the following how all quark masses and mixing angles are 
realistically obtained in this model.

We note first that our quark mass matrices are restricted by our choice 
of $A_4$ representations to have only 5 independent parameters each. 
In the $down$ sector, the Yukawa couplings $h_{1,2,3}$ can all be chosen 
real, $A$ is just an overall scale, and $B$ is complex. 
The 5 independent parameters can be chosen as the 3 quark masses, and 
2 angles.  In the $up$ sector, we can choose $\mu_2$ and $m_{2,3}$ to be 
real, with $\mu_3$ complex.  The 5 independent parameters can be chosen 
as the 3 quark masses, 1 angle, and 1 phase.  Since we also have 10 
observables (6 quark masses, 3 angles, and 1 phase), it may appear that 
a fit is not so remarkable.  However, the forms of the 2 mass matrices 
are very restrictive, and it is by no means trivial to obtain a good fit. 
Indeed, we find that $V_{ub}$ is strongly correlated with the CP phase 
 $\beta$.  If we were to fit just the 6 masses and the 3 angles, the 
structure of our mass matrices would allow only a very 
narrow range of values for $\beta$ at each value of $|V_{ub}|$.  This 
means that future more precise determinations of these two parameters 
will be a decisive test of this model.

Our quark mass matrices are given at the SU(5) unification scale in principle.
However, the $A_4$ flavor symmetry is spontaneously broken at the electroweak
scale. Therefore, the forms of our mass matrices are not changed except for 
the magnitudes of the Yukawa couplings between the unification and 
electroweak scales.  Our numerical analyses are presented at the 
electroweak scale.

In order to fit the ten observables (six quark masses, three CKM mixing 
angles and one phase), 1,000,000 
random numbers have been generated for the ten parameters of our model.  
We then choose the parameter sets which are allowed by the experimental
data. First we show the prediction of $|V_{ub}|$ versus $\beta$ in Figure~1,
 with the following nine experimental inputs \cite{mass,PDG,CP}:
\begin{eqnarray}
 && m_u = 0.9 \sim 2.9 \ {\rm (MeV)},\qquad
 m_c = 530 \sim 680 \ {\rm (MeV)}, \qquad
 m_t = 168 \sim 180 \ {\rm (GeV)}, 
\nonumber \\
 && m_d = 1.8 \sim 5.3\ {\rm (MeV)}, \qquad
 m_s = 35 \sim 100 \ {\rm (MeV)}, \qquad
 m_b = 2.8 \sim 3 \ {\rm (GeV)},   \label{input1} \\
&& |V_{us}| = 0.221 \sim 0.227,\qquad
 |V_{cb}| = 0.039 \sim 0.044,\qquad
 J_{CP} = (2.75 \sim 3.35)\times 10^{-5}, \nonumber
\end{eqnarray}
\noindent
which are given at the electroweak scale.
Here $J_{CP}$ is the Jarlskog invariant \cite{Jarlskog}.
\begin{figure}
\begin{center}
\includegraphics[scale=1.]{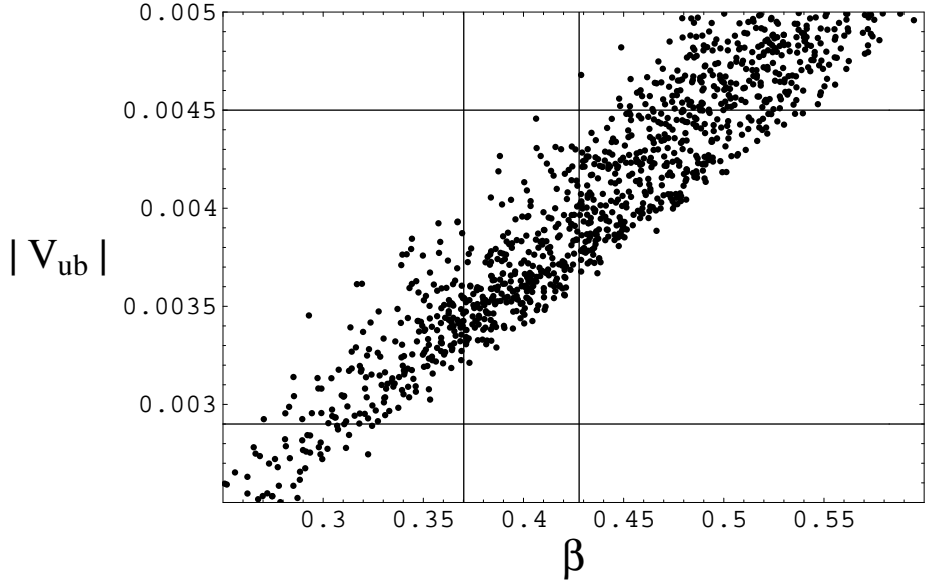}
\end{center}
\caption{Plot of allowed values  in the $\beta - |V_{ub}|$ plane, where
the value of $\beta$ is expressed in radians.
 The horizontal and vertical lines denote experimental bounds at $90\%$ C.L.}
\end{figure}
We see that the experimental allowed region of $\beta$ 
($0.370\sim 0.427$ radian  at $90\%$ C.L.) \cite{CP} 
corresponds to $|V_{ub}|$ in the range $0.0032\sim 0.0044$,
which is consistent with the experimental value
of $|V_{ub}|=0.0029\sim 0.0045$.  Thus our model is able to reproduce 
realistically the experimental data of 
quark masses and the CKM matrix.

Precisely measured heavy quark masses and CKM matrix elements are expected 
in future experiments and precise light quark masses are expected in 
future lattice evaluations. 
If the allowed regions of the current data shown 
in Eq.~(\ref{input1}) are reduced,
the correlation between  $|V_{ub}|$ and $\beta$ will become stronger. 
We show in Figure~2 the case where the experimental data are 
restricted to some very narrow ranges about their central values: 
\begin{eqnarray}
 && m_u/m_d = 0.5\sim 0.6,\qquad\ \ \
 m_c = 600 \sim 610 \ {\rm (MeV)}, \quad
 m_t = 172 \sim 176 \ {\rm (GeV)}, 
\nonumber \\
 && m_d = 3.4 \sim 3.6\ {\rm (MeV)}, \quad\ \ 
 m_s/m_d = 18\sim 20, \qquad\quad\ 
 m_b = 2.85 \sim 2.95 \ {\rm (GeV)}, \label{input2}\\
&& |V_{us}| = 0.221 \sim 0.227,\qquad
 |V_{cb}| = 0.041 \sim 0.042,\qquad\ 
  J_{CP} = (3.0 \sim 3.1) \times 10^{-5}. \nonumber
\end{eqnarray}
Here we use the tighter constraints on the mass ratios of light quarks, 
i.e. $m_u/m_d$ and $m_s/m_d$, consistent with the well-known successful 
low-energy sum rules \cite{lightmass}.
Clearly, future more precise determinations of $|V_{ub}|$ and $\beta$ 
will be a sensitive test of our model.
\begin{figure}
\begin{center}
\includegraphics[scale=1]{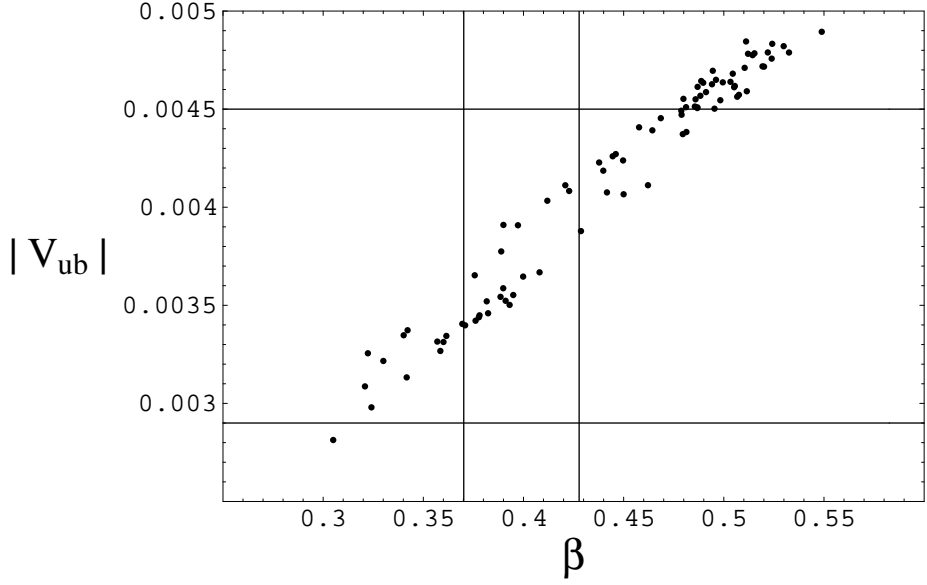}
\end{center}
\caption{Plot of allowed values  in the $\beta - |V_{ub}|$ plane, where
input data are restricted in the narrower regions shown in Eq.~(\ref{input2}).}
\end{figure}

 We should also comment on  the hierarchy of $h_i$ and $v_i$.
 The order of $h_i$ are fixed by  quark mixings.
The ratio of $h_2/h_3\simeq \lambda^2$  is required by the $V_{cb}$
mixing  ($\lambda\simeq 0.22$), on the other hand,  
$h_1/h_2\simeq \lambda$ comes from $V_{us}$. 
 Once $h_i$ are fixed, quark masses determine
 the hierarchy of $v_i$ as follows:
 $v_1/v_3\simeq \lambda^2$ and
  $v_2/v_3\simeq \lambda \sim \lambda^{1/2}$.
These hierarchies of $h_i$ and $v_i$
are also consistent with the magnitude of $J_{CP}$,
which is given by 
\begin{equation}
 J_{CP}\simeq \frac{\sqrt{3}}{2}\frac{h_1^2}{h_3^2}
\left( \frac{v_2^2 - v_1^2}{v_2^2 + v_1^2} \right)
\left (1+ \frac{{\rm Re}(\mu_3)+\frac{1}{\sqrt{3}}{\rm Im}(\mu_3)}{m_t}
\frac{h_3}{h_1} \right ) \ .
\end{equation}

In summary, the $A_4$ family symmetry (which has been successful in 
understanding the mixing pattern of neutrinos) is applied successfully 
as well to quarks, motivated by the quark-lepton assignments of SU(5).  
The Yukawa interactions of this model are invariant under $A_4$, but the 
vacuum expectation values of the Higgs scalars are allowed to be arbitrary. 
The resulting $u$ and $d$ quark mass matrices have 5 parameters each, 
with different specific structures. Realistic quark masses and mixing angles 
are obtained, in good agreement with data.  In particular, we find a 
strong correlation between $|V_{ub}|$ and the CP phase $\beta$, thus 
allowing for a decisive  future test of this model.

\vskip 1 cm

This work was supported in part by the U.~S.~Department of Energy under
Grant No. DE-FG03-94ER40837
 and  the Grant-in-Aid for Science Research,
 Ministry of Education, Science and Culture, Japan (No.~17540243).
  EM thanks the Department of Physics,  
Niigata University for hospitality during a recent visit.
     
\newpage
\bibliographystyle{unsrt}

\begin{thebibliography}{99}
\bibitem{mr01} E. Ma and G. Rajasekaran, Phys. Rev. {\bf D64}, 113012 (2001). 
\bibitem{bmv03} K. S. Babu, E. Ma, and J. W. F. Valle, Phys. Lett. 
{\bf B552}, 207 (2003).
\bibitem{hps02} P. F. Harrison, D. H. Perkins, and W. G. Scott, Phys. 
Lett. {\bf B530}, 167 (2002).  See also X.-G. He and A. Zee, Phys. Lett. 
{\bf B560}, 87 (2003).
\bibitem{a432} See for example E. Ma, Phys. Rev. {\bf D70}, 031901R (2004); 
Phys. Rev. {\bf D72}, 037301 (2005); {\bf D73}, 057304 (2006); 
G. Altarelli and F. Feruglio, Nucl. Phys. {\bf B720}, 64 (2005); 
{\bf B741}, 215 (2006); K. S. Babu and X.-G. He, hep-ph/0507217.
\bibitem{bdm99} K. S. Babu, B. Dutta, and R. N. Mohapatra, Phys. Rev. 
{\bf D60}, 095004 (1999).
\bibitem{a4q} E. Ma, Mod. Phys. Lett. {\bf A17}, 627 (2002); X.-G. He, 
Y.-Y. Keum, and R. R. Volkas, JHEP {\bf 0604}, 039 (2006).
\bibitem{fuji} E. Ma, hep-ph/0409075.
\bibitem{mass} H. Fritzsch and Z. Xing,
Prog. Part. Nucl. Phys. {\bf 45}, 1 (2000). 
\bibitem{PDG}
Particle Data Group,  http://pdg.lbl.gov/ (2006).
\bibitem{CP}  J. Charles (CKM fitter group),  hep-ph/0606046.
\bibitem{Jarlskog} C. Jarlskog, Phys. Rev. Lett. {\bf 55}, 1039 (1985).
\bibitem{lightmass}  H. Leutwyler,  
Phys. Lett. {\bf B378}, 313 (1996), hep-ph/9609467.
\end{thebibliography}

\end{document}